\documentclass[runningheads]{llncs}
\usepackage{graphicx}
\usepackage{tikz}
\usepackage{epstopdf}
\usepackage{pgfplots}
\usepackage{fancybox}
\usepackage{lipsum}
\usepackage[many]{tcolorbox}    
\usepackage{wrapfig}
\usepackage{calc}
\usepackage{multicol}
\usepackage{rotating}

\begin{document}
\title{Conversational Agents for Insurance Companies - From Theory to Practice}
%
%
\author{Falko Koetter\inst{1} \and
Matthias Blohm\inst{1} \and
Jens Drawehn\inst{1} \and
Monika Kochanowski\inst{1} \and
Joscha Goetzer\inst{2} \and
Daniel Graziotin\inst{2} \and
Stefan Wagner\inst{2}}
\authorrunning{F. Koetter et al.}
%
\institute{Fraunhofer Institute for Industrial Engineering, Nobelstr. 12, 70569 Stuttgart, Germany\\
\email{falko.koetter@iao.fraunhofer.de, matthias.blohm@iao.fraunhofer.de, jens.drawehn@iao.fraunhofer.de, monika.kochanowski@iao.fraunhofer.de}\\
 \and
University of Stuttgart, Universit\"atsstr. 38, 70569 Stuttgart, Germany\\
\email{joscha.goetzer@gmail.com, daniel.graziotin@iste.uni-stuttgart.de, stefan.wagner@iste.uni-stuttgart.de}
}
\maketitle              
\begin{abstract}
Advances in artificial intelligence have renewed interest in conversational agents. Additionally to software developers, today all kinds of employees show interest in new technologies and their possible applications for customers. German insurance companies generally are interested in improving their customer service and digitizing their business processes. In this work we investigate the potential use of conversational agents in insurance companies theoretically by determining which classes of agents exist which are of interest to insurance companies, finding relevant use cases and requirements. We add two practical parts: First we develop a showcase prototype for an exemplary insurance scenario in claim management. Additionally in a second step, we create a prototype focusing on customer service in a chatbot hackathon, fostering innovation in interdisciplinary teams. In this work, we describe the results of both prototypes in detail. We evaluate both chatbots defining criteria for both settings in detail and compare the results and draw conclusions for the maturity of chatbot technology for practical use, describing the opportunities and challenges companies, especially small and medium enterprises, face. 

\keywords{conversational agents \and intelligent user interfaces \and hackathon \and nlp \and chatbot \and insurance.
}
\end{abstract}
\section{\uppercase{Introduction}}

\label{sec:introduction}

With the digital transformation changing usage patterns and consumer expectations, many industries need to adapt to new realities. The insurance sector is next in line to grapple with the risks and opportunities of emerging technologies, in particular \emph{Artificial Intelligence}~\cite{aichanging}. Additionally, innovation methods like design thinking and open innovation are on the rise. In unsecure market times innovation is crucial, and all organizations and also traditional companies need to keep up to date by using new technologies for innovative business processes \cite{soltani14}.

Fraunhofer~IAO as an applied research institution supports digital transformation processes in an ongoing project with multiple insurance companies~\cite{iao2018innonetz}. The goal of this project is to scout new technologies, investigate them, rate their relevance and evaluate them (e.g. in a model trial or by implementing a prototype). While insurance has traditionally been an industry with very low customer engagement, insurers now face a young generation of consumers with changing attitudes regarding insurance products and services~\cite{iao2017zukunftsstudie}. Another goal of the project is the establishment of innovation methods within the companies and enable them to develop new products and services themselves.

Traditionally, customer engagement uses channels like mail, telephone and local agents. In 2016, chatbots emerged as a new trend~\cite{customerservice}, making it a topic of interest for Fraunhofer~IAO and insurance companies. With the rise of the smartphone, many insurers started offering apps, but success was limited~\cite{jdpower2017}, which may stem from app fatigue~\cite{appfatigue}. App use has plateaued, as users have too many apps and are reluctant to add more~\cite{gartner}. In contrast, conversational~agents require no separate installation, as they are accessible via messaging apps, which are likely to be already installed on a user's smartphone. Conversational agents are an alternative to improve customer support and digitize processes like claim handling or managing customer data.

The objective of this work is to describe the creation of conversational agents in theory and practice and show the outcomes of both views. We facilitate the creation of conversational agents by defining the traits of an agent more clearly using a (1) classification framework, which is based on current literature and research topics, and systematically analyzing (2) use cases and requirements in an industry, shown in the example insurance scenario. We frame two application scenarios with this theoretical foundation. Prototype 1 is a claim-handling scenario, which shows technological progress for a conversational agent. In this extended version of our former paper~\cite{koetter18}, we present prototype 2. This new prototype has been created for the scenario of customer service and cross selling. It is created in the setting of a chatbot hackathon event that Fraunhofer IAO organized in 2018. The goal is to gain more insights about conversational agent creation while examining the practicability of chatbot implementation for small insurance scenarios. Furthermore, we enriched the evaluation chapter of both prototypes and compare the results of both activities. We derive possible applications, knowledge about challenges and success factors as learnings from both activities. We apply this knowledge in a new project for supporting small and medium enterprises in adoption of new technologies. 

\section{\uppercase{Related Work}}
\label{sec:relatedwork}

In this section we investigate work in the area of conversational agents, dialog management, and research applications in insurance. In extension to the previous paper \cite{koetter18}, we add theory on hackathons at the end of the section.

\cite{theconversationalinterface} offer detailed explanations about background and history of conversational interfaces as well as techniques to build and evaluate own agent applications. Another literature review about chatbots was provided by \cite{chatbotthesis}, where common approaches and design choices are summarized followed by a case study about the functioning of IBM's chatbot Watson, which became famous for winning the popular quiz game \emph{Jeopardy!} against humans.

Many chatbot applications have already been built nowadays with the goal to solve actual problems. One example is PriBot, a conversational agent, which can be asked questions about an application's privacy policy, because users tended to skip reading the often long and difficult to understand privacy notices. Also, the chatbot accepts queries of the user which aim to change his privacy settings or app permissions~\cite{pribots}.

In the past there have already been several studies with the goal to evaluate how a conversational agent should behave for being considered as human-like as possible. In one of them, conducted by~\cite{hallmarks}, fourteen participants were asked to talk to an existing chatbot and to collect key points of convincing and unconvincing characteristics. It turned out that the bot's ability to hold a theme over a longer dialog made it more realistic. On the other hand, not being able to answer to a user's questions was regarded as an unsatisfying characteristic of the artificial conversational partner~\cite{hallmarks}.

In another experiment, which was done by ~\cite{onboarding}, eight users had to talk to two different kinds of chatbots, one behaving more human-like and one behaving more robotic. In this context, they had to fulfill certain tasks like ordering an insurance policy or demanding an insurance certification. All of the participants instinctively started to chat by using natural human language. In cases in which the bot did not respond to their queries in a satisfying way, the users' sentences continuously got shorter until they ended up with writing key words only. Thus, according to the results of this survey, conversational agents preferably should be created human-like, because users seem to be more comfortable when feeling like talking to another human being, especially in cases in which the concerns are crucial topics like their insurance policies~\cite{onboarding}. 

\emph{Dialog management strategies} (DM) define the conversational behaviors of a system in response to user message and system state~\cite{theconversationalinterface}. 

In industry applications, DM often consists of a handcrafted set of rules and heuristics, which are tightly coupled to the application domain \cite{theconversationalinterface} and improved iteratively. One problem with handcrafted approaches to DM is that it is challenging to anticipate every possible user input and react appropriately, making development resource-intensive and error-prone. But if few or no recordings of conversations are available, these \emph{rule-oriented} strategies may be the only option.

As opposed to the rule-oriented strategies, data-oriented architectures work by using machine learning algorithms that are trained with samples of dialogs in order to reproduce the interactions that are observed in the training data. 
These statistical or heuristical approaches to DM can be classified into three main categories: Dialog modeling based on \emph{reinforcement learning}, \emph{corpus-based} statistical dialog management, and \emph{example-based} dialog management (simply extracting rules from data instead of manually coding them)~\cite{theconversationalinterface}\cite{spierling2005interactive}. \cite{spierling2005interactive} highlights neural networks, Hidden-Markov Models, and  Partially Observable Markov Decision Processes as possible implementation technologies.

The following are common strategies for rule-based dialog management:

\begin{itemize}

\item Finite-state-based DM uses a finite state machine with handcrafted rules, and performs well for highly structured, system-directed tasks~\cite{theconversationalinterface}.
\item Frame-based DM follows no predefined dialog path, but instead allows to gather pieces of information in a frame structure and no specific order. This is done by adding an additional entity-value slot for every piece of information to be collected and by annotating the intents in which they might occur. Using frames, a less restricted, user-directed conversation flow is possible, as data is captured as it comes to the mind of the user~\cite{rudnicky1999agenda}.
\item Information State Update represents the information known at a given state in a dialog and updates the internal model each time a participant performs a \emph{dialog move}, (e.g. asking or answering). The state includes information about the mental states
of the participants (beliefs, desires, intentions, etc.) and about the dialog (utterances, shared information, etc.) in abstract representations. Using so-called update moves, applicable moves are chosen based on the state~\cite{traum2003isu}.
\item Agent-based DM uses an  agent that fulfills conversation goals by dynamically using plans for tasks like intent detection and answer generation. The agent has a set of beliefs and goals as well as an information base which is updated throughout the conversation. Within this information framework the agent continuously prioritizes goals and autonomously selects plans that maximize the likelihood of goal fulfillment~\cite{nguyen2005agent}.

\end{itemize}

\cite{agentdm} describes how multiple DM approaches can be combined to use the best strategy for specific circumstances.

A virtual insurance conversational agent is described by \cite{yacoubi2018teatime}, utilizing \emph{TEATIME}, an architecture for agent-based DM. TEATIME uses emotional state as a driver for actions, e.g. when the bot is perceived unhelpful, that emotion leads the bot to apologize. The shown example bot is a proof of concept for TEATIME capable of answering questions regarding insurance and react to customer emotions, but does not implement a full business process.

\cite{kowatsch2017text} describe a text-based healthcare chatbot that acts as a companion for weightloss but also connects a patient with healthcare professionals. The chat interface supports non-textual inputs like scales and pictorials to gather patient feedback. Study results showed a high engagement with the chatbot as a peer and a higher percentage of automated conversation the longer the chatbot is used.

Overall, these examples show potential for conversational agents in the insurance area, but lack support for complete business processes.

Considering \emph{hackathons} previous research has been done on (examples include \cite{soltani14} and \cite{briscoe2014hackathon}). Important for hackathons are \emph{goals} of a hackathon as well as \emph{success factors}. Hackathons are problem-focused computer programming events in which teams of programmers and other stakeholders prototype a software solution within a limited timeframe~\cite{briscoe2014hackathon}. Hackathons usually are characterized by three features: (1) intensive collaborative work experience (2) solution of a concrete problem with a demonstrable solution (3) and a short time span. Depending on the focus and target group several specific formats are possible, like internal or external or application or technology specific hackathons\cite{briscoe2014hackathon}. Much work apart from the work cited here on hackathons has been published. To the best of our knowledge, an internal but company-spanning conversational agent hackathon in the insurance industry has not been described yet. We will compare the resulting prototypes based on the same technology of the hackathon with the prototype developed within a traditional project setting for deriving potentials and success factors for conversational agent creation. Furthermore, we will compare technological progress of the resulting prototypes.

\section{\uppercase{Theory on Conversational Agents and Insurance Industry}}
\label{sec:agents}
\subsection{Application Scenarios and Types of Agents}
\label{subsec:application}

The idea of conversational agents that are able to communicate with human beings is not new: In 1966, Joseph Weizenbaum introduced \emph{Eliza}, a virtual psychotherapist, which was able to respond to user queries using natural language and which could be considered as the first \emph{chatbot}~\cite{eliza1966}. However, Eliza used quite simple structures by just picking up keywords and asking more questions, not serving a purpose itself. Nowadays, the idea of speaking machines has experienced a revival with the emergence of new technologies, especially in the area of artificial intelligence. Novel machine learning algorithms allow developers to create software agents in a much more sophisticated way and in many cases they already outperform previous statistical NLP methods~\cite{theconversationalinterface}. Additionally, the importance of messaging apps such as WhatsApp or Telegram has increased over the last years. In 2015, the total number of people using these messaging services outran the total number of active users in social networks for the first time. Today, each of these app has about between 200 million and 1.5 billion users~\cite{messengers}. Currently the topic voice is on the rise - not only Gartner considers the breakthrough of voice applications in the next years.

Conversational agents can be basically employed in these settings: 
\begin{itemize}
	\item \textbf{Customer service} In 2016~\cite{customerservice} the topic of customer service chatbots lead to a great variety with a wide range of terminology. 
	\item \textbf{Recruitment} Recruitment chatbots become more popular, also the insurance company Allianz launched a recruitment bot recently.
	\item \textbf{Marketing} Chatbots can be used for giving a company an innovative and up-to-date view without really serving a business process.
	\item \textbf{Internal support} Before chatbots became so popular, many companies already used chatbots for internal purposes. One example is IBM with its' "Whatis Bot"\footnote{https://www.academia.edu/35150361/IBM\_whatis}, which answered questions by instant messaging about acronyms already many years ago. The requirements for internal chatbots tend to be lower than for external ones, as customers usually have the choice of a communication channel or provider. 
\end{itemize}

This paper focuses on \emph{customer service} only for demonstration purposes and simple explainability. However, in the insurance project mentioned beforehand, the second category of internal support by NLP chatbots or voice systems has gathered even more interest. 

For being able to draw a big picture of the current trends in the area of conversational agents, we divide them into the following four common categories:

\begin{itemize}
	\item \textbf{(Virtual, Intelligent, Cognitive, Digital, Personal) assistants
  (VPAs)}: Agents fulfilling tasks intelligently based on spoken or written user input and with the help of data bases and personalized user preferences~\cite{cooper2008personal} (e.g. Apple's Siri or Amazon's Alexa~\cite{chatbotsreturn}).
	\item \textbf{Specialized digital assistants (SDAs)}: Focused on a specific domain of expertise, goal-oriented behavior ~\cite{chatbotsreturn}. SDAs can be used in customer service as well as for internal support tasks.
	\item \textbf{Embodied conversational agents (ECAs)}: Visually animated agents, e.g. in form of avatars or robots~\cite{evaluating2017}, where speech is combined with gestures and facial expressions.
	\item \textbf{Chatterbots}: Bots with focus on small talk and realistic conversations, not task-oriented, e.g. Cleverbot~\cite{cleverbot}.
\end{itemize}

\begin{figure}[htbp]
	\usetikzlibrary{shapes}

\newcommand{\DU}{6} 
\newcommand{\UU}{7} 

\newdimen\R 
\R=3.5cm 
\newdimen\L 
\L=4.1cm

\newcommand{\AW}{360/\DU} 



	

    \begin{tikzpicture}[scale=0.5]
      \path (0:0cm) coordinate (O); 
			
      \foreach \X in {1,...,\DU}{
        \draw (\X*\AW:0) -- (\X*\AW:\R);
        \draw (\X*\AW:0) -- (\X*\AW:\R);
      }

      \foreach \Y in {0,...,\UU}{
        \foreach \X in {1,...,\DU}{
          \path (\X*\AW:\Y*\R/\UU) coordinate (D\X-\Y);
          \fill (D\X-\Y) circle (1pt);
        }
        \draw [opacity=0.3] (0:\Y*\R/\UU) \foreach \X in {1,...,\DU}{
            -- (\X*\AW:\Y*\R/\UU)
        } -- cycle;
      }

      \path (1*\AW:\L) node (L1) {\footnotesize \textit{\textsf{Companionship}}};
      \path (2*\AW:\L) node (L2) {\footnotesize \textit{\textsf{Realism}}};
      \path (3*\AW:\L) node (L3) {\footnotesize \textit{\textsf{Entertainment}}~~~~~~~~~~~~~~~~~};
      \path (4*\AW:\L) node (L4) {\footnotesize \textit{\textsf{Textual}}};
      \path (5*\AW:\L) node (L5) {\footnotesize \textit{\textsf{Task Orientation}}};
      \path (6*\AW:\L) node (L6) {\footnotesize ~~~~~\textit{\textsf{Spoken}}};

      \newcommand{\bgc}{null}
      %
      %
      \newcommand{\opazetee}{0.35}
      \newcommand{\linewidthall}{2.1pt}

      \renewcommand{\bgc}{magenta}
      \draw [color=\bgc,line width=\linewidthall,opacity=\opazetee,fill=\bgc]
        (D1-4) --
        (D2-6) --
        (D3-7) --
        (D4-7) --
        (D5-1) --
        (D6-2) -- cycle;
        \node[text width=5cm,color=\bgc] at (-1.8,-0.8) 
            {\scriptsize \textbf{Chatterbots}};

      \renewcommand{\bgc}{green}
      \draw [color=\bgc,line width=\linewidthall,opacity=\opazetee,fill=\bgc]
        (D1-3) -- 
        (D2-2) --
        (D3-3) --
        (D4-2) --
        (D5-6) --
        (D6-7) -- cycle;
        \node[text width=5cm,color=\bgc] at (8.55,-1.0) 
            {\scriptsize \textbf{General\\Digital Assistants}};

      \renewcommand{\bgc}{violet}
      \draw [color=\bgc,line width=\linewidthall,opacity=\opazetee,fill=\bgc]
        (D1-7) --
        (D2-7) --
        (D3-4) --
        (D4-1) --
        (D5-1) --
        (D6-5) -- cycle;
        \node[text width=5cm,color=\bgc] at (7.55,2.3) 
            {\scriptsize \textbf{Embodied\\Conversational Agents}};

      \renewcommand{\bgc}{yellow}
      \draw [color=\bgc,line width=\linewidthall,opacity=\opazetee,fill=\bgc]
        (D1-1) --
        (D2-1) --
        (D3-2) --
        (D4-6) --
        (D5-7) --
        (D6-4) -- cycle;
        \node[text width=5cm,color=\bgc] at (7.80,-2.55) 
            {\scriptsize \textbf{Specialized\\Digital Assistants}};

      \renewcommand{\bgc}{black}
      \draw [color=\bgc,line width=\linewidthall,opacity=\opazetee,fill=\bgc]
        (D1-1) --
        (D2-1) --
        (D3-4) --
        (D4-4) --
        (D5-5) --
        (D6-3) -- cycle;
        \node[text width=5cm,color=\bgc] at (7.80,-4.55) 
            {\scriptsize \textbf{*Prototypes developed}};

       \renewcommand{\bgc}{black}
			\node[color=blue,outer sep=1] at (D1-0) {\textbf{\small 0}};
			\node[color=blue,outer sep=1] at (D1-1) {\textbf{\small 1}};
			\node[color=blue,outer sep=1] at (D1-2) {\textbf{\small 2}};
			\node[color=blue,outer sep=1] at (D1-3) {\textbf{\small 3}};
			\node[color=blue,outer sep=1] at (D1-4) {\textbf{\small 4}};
			\node[color=blue,outer sep=1] at (D1-5) {\textbf{\small 5}};
			\node[color=blue,outer sep=1] at (D1-6) {\textbf{\small 6}};
			\node[color=blue,outer sep=1] at (D1-7) {\textbf{\small 7}};


      \end{tikzpicture}
	\centering
       \caption{Classification of conversational agents with their characteristics (based on own presentation from \cite{koetter18}). Values between 0 and 7 indicate how strong a characteristic applies for the given type of agent. Additionally to the classification, the prototype implementations are shown by the black box.}\label{fig:cui-types}
\end{figure}
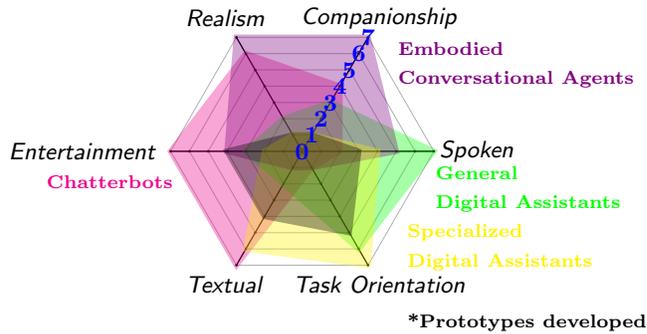

Figure \ref{fig:cui-types} shows the results of evaluating these four classes in terms of different characteristics such as \emph{realism} or \emph{task orientation} based on own literature research. Chatterbots provide a high degree of entertainment since they try to imitate the behavior of human beings while chatting, but there is no specific goal to be reached within the scope of these conversations. In contrast, general assistants like Siri or Alexa are usually called by voice in order to fulfill a specific task. Specialized assistants concentrate even more on achieving a specific goal, which often comes at the expense of realism and user amusement because their ability to respond to not goal-oriented conversational inputs like small talk is mostly limited. The best feeling of companionship can be experienced by talking to an embodied agent, since the reactions of these bots are closest to human-like behavior.

Taking a look at the insurance project, it was decided to create prototypes for customer service in the type of \emph{specialized digital assistants}. In the next paragraph, the processes in the insurance domain which might be chosen for this implementation are described. As shown in figure \ref{fig:cui-types}, it has shown that although the goal was to created a specialized digital assistant, humans have their own goals in prototype creation. Adding small talk in a limited scope affected the prototype creation and led to a more realistic and human-like user experience and more entertainment for the prototype in the hackathon as well.

\subsection{Insurance Processes and Requirements for Prototypes}
\label{subsec:insurance}

Insurance is an important industry sector in Germany, with 560 companies that manage about 460 million policies~\cite{gdv2014statistik}. However, the insurance sector is under a high cost pressure, which shows in a declining employee count and low margins~\cite{stange2015zukunft}. The insurance market is saturated and has transitioned from a growth market to a displacement market~\cite{aschenbrenner2010versicherung}. For the greater part, German insurance companies have used conservative strategies, caused by risk aversion, long-lived products, hierarchical structures, and profitable capital markets~\cite{veraenderungsaversion}. As these conditions change, so must insurance companies. One effort is the insurance project ~\cite{iao2018innonetz} with the goal of innovation and new technologies performed by Fraunhofer IAO since several years as described in section \ref{sec:introduction}. The two touch points of interest in the insurance industry are \emph{selling a product} and \emph{the claims process}. A study found that consumers interact less with insurers than with any other industry ~\cite{engagement}. This is the reason why although chatbots become more popular in other use cases like recruitment, the focus of this paper is the application in customer service.

Many insurance companies have heterogeneous IT infrastructures incorporating legacy systems (sometimes from two or more companies as the result of a merger)~\cite{digitalenterprise}. These grown architectures pose challenges when implementing new data-driven or AI solutions, due to issues like data quality, availability and privacy. Nonetheless, the high amount of available data and complex processes make insurance a prime candidate for machine learning and data mining. The adoption of AI in the insurance sector is in early stages, but accelerating, as insurance companies strive to improve service and remain competitive~\cite{aichanging}.

Conversational agents are one AI technology at the verge of adoption. In 2017, ARAG launched a travel insurance chatbot, quickly followed by bots from other insurance companies~\cite{aragchatbot}. Examples are a chatbot on moped insurance by wgv\footnote{https://www.wgv.de/versicherungen/kfz/moped/} and a chatbot on car insurance by Allianz\footnote{https://www.facebook.com/AllianzCarlo/}. 

\begin{figure}[htbp]
	\includegraphics[width=0.8\textwidth]{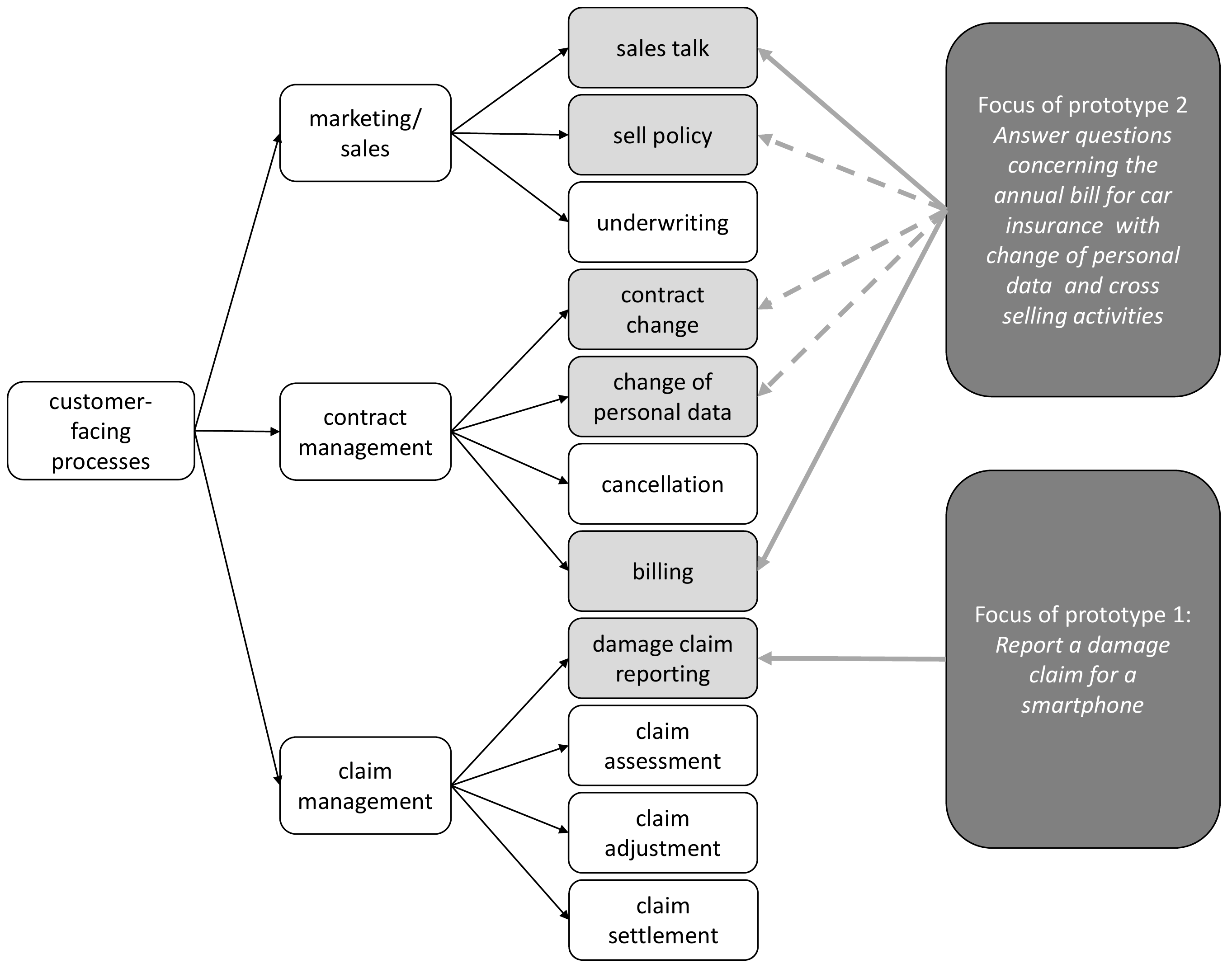}
	\centering
	\caption{Customer-facing insurance processes (original in \cite{koetter18} based on~\cite{aschenbrenner2010versicherung} and \cite{horch2012openxchange}) with additional information on the prototypes as described in this paper shown in grey}
	\label{fig:processes}
\end{figure}

To identify areas of possible chatbot support, we surveyed the core business processes of insurance companies as described in ~\cite{aschenbrenner2010versicherung} and \cite{horch2012openxchange}. Three core areas of insurance companies are customer-facing: \emph{marketing/sales}, \emph{contract management} and \emph{claim management}. Figure~\ref{fig:processes} shows the main identified processes related to this area. 

We identified all these processes as possible use cases for conversational agent support, in particular support by SDAs. As two prototypes are planned, the criteria are analyzed for both settings. The chosen scenario for prototype 1 is a special case of the damage claim process: \emph{The user has a damaged smartphone or tablet and wants to make an insurance claim}. The scenario for prototype 2 in the hackathon is: \emph{The user has received an annual bill. Answer frequently asked questions concerning the annual bill for car insurance and combine with change of personal data and cross selling activities} (see also figure \ref{fig:processes}). 

Furthermore, we investigated general requirements for conversational agents in these processes:

\textbf{Availability and ease-of-use} Conversational agents are an alternative to both conventional customer support (e.g. phone, mail) as well as conventional applications (e.g. apps and websites). Compared to these conventional solutions, chatbots offer more availability than human agents and have less barriers of use than conventional applications, requiring neither an installation nor the ability to learn a new user interface, as conventional messaging services are used~\cite{derler2017}. This includes the requirement of understanding and answering to human language, which applies to both prototypes developed.

\textbf{Guided information flow} Compared to websites, which offer users a large amount of information they must filter and prioritize themselves, conversational agents offer information gradually and only after the intent of the user is known. Thus, the search space is narrowed at the beginning of the conversation without the user needing to be aware of all existing options. This is done for both prototypes by narrowing the scope.

\textbf{Smartphone integration} Using messaging services, conversational agents can integrate with other smartphone capabilities, e.g. making a picture, sending a calendar event, setting a reminder or calling a phone number. This applies for both prototypes.

\textbf{Customer call reduction} Customer service functions can be measured by reduction of customer calls and average handling time~\cite{customerservice}. SDAs can help here by automating conversations, handling standard customer requests and performing parts of conversations (e.g. authentication). This is relevant for projects, but out of scope for the prototype. However, questions about the annual bill arise very frequently.

\textbf{Human handover} Customers often use social media channels to escalate an issue in the expectation of a \emph{human} response, instead of an \emph{automated} one. A conversational agent thus must be able to differentiate between standard use cases it can handle and more complicated issues, which need to be handed over to human agents~\cite{risk}. One possible approach is to use sentiment detection, so customer who are already stressed are not further aggravated by a bot~\cite{customerservice}. Being out of scope for the prototype, this has only be investigated for some technology providers that have different levels of experience with this question.

\textbf{Digitize claim handling} Damage claim handling in insurance companies is a complex process involving multiple departments and stakeholders~\cite{koetter2012business}. Claim handling processes are more and more digitized within the insurance companies~\cite{horch2012openxchange}, but paper still dominates communication with claimants, workshops and experts. \cite{touchless} defines maturity levels of insurance processes, defining \emph{virtual handling} as a process where claims are assessed fully digitally based on digital data from the claimant (e.g. a video, a filled digital form), and \emph{touchless handling} as a fully digital process with no human intervention on the insurance side. SDAs help moving towards these maturity levels by providing a guided way to make a claim digitally and communicate with the claimant (e.g. in case additional data is needed). Prototype 1 covers this area. 

\textbf{Conversational commerce} is the use of Conversational Agents for marketing and sales related purposes~\cite{commerce}. Conversational Agents can perform multiple tasks using a single interface. Examples are using opportunities to sell additional products (\emph{cross-sell}) or better versions of the product the customer already has (\emph{up-sell}) by chiming in with personalized product recommendations in the most appropriate situations. One example would be to note that a person's last name has changed during an address update customer service case and offer appropriate products if the customer has just married. Prototype 2 covers this area. 

\textbf{Internationalization} is an important topic for large international insurance companies. However, most frameworks for implementing conversational agents are available in more than one language. To the best of our knowledge, the applied conversational agents in German insurance today are optimized only for one language. So this topic is future work in respect to both prototypes, but will become more important in the future.

\textbf{Compliance} to privacy (GDPR) is usually guaranteed by the login mechanisms on the insurance sites, therefore the topic is out of scope for our research prototype. For broader scenarios not requiring identification on the insurance site and the usage of the data for non-costumers, this is an area of ongoing research on compliant technical solutions or workarounds.

\section{\uppercase{Practice in two Prototypes in Insurance Industry}}
\label{sec:prototype}

\subsection{Technical requirements and framework options for the prototypes}
\label{subsec:requirements}

For dialog design within prototype 1, experimenting with machine learning algorithms was the preferred implementation strategy. For this purpose, discussions with insurance companies were held to assess the feasibility of receiving existing dialogs with customers, for example for online chats, phone logs or similar. However, such logs generally seem to be not available at German insurers, as the industry has self-regulated to only store data needed for claim processing~\cite{wwwGDVCOC2012}. As a research institute represents a third party not directly involved in claims processing, data protection laws forbid sharing of data this way without steps to secure personal data. During our talks we have identified a need for automated or assisted anonymization of written texts as a precondition for most customer-facing machine learning use cases, at least when operating in Europe~\cite{kamarinou2016machine}. However, these issues go beyond the scope of our current project, but provide many opportunities for future research. 

To still build a demonstrator in face of these challenges as outlined in \cite{koetter19}, dialogs for both prototypes were manually designed without using real-life customer conversations and fine-tuned by user testing with fictional issues. As this approach entails higher manual effort for dialog design, a narrower scenario was chosen for both prototypes to still allow for the full realization of a customer-facing process. 

Based on the work presented in the last sections and our talks with insurance companies, we arrived at the following non-functional requirements that the chatbot prototype 1 ideally should fulfill: 

\begin{itemize}
	\item \textbf{Interoperability}: The agent should be able to keep track of the conversational context over several message steps and messengers.
	\item \textbf{Portability}: The agent can be run on different devices and platforms (e.g. Facebook Messenger, Telegram). Therefore it should use a unified, platform-independent messaging format.
	\item \textbf{Extensibility}: The agent should provide a high level of abstraction that allows designers to add new conversational content without having to deal with complicated data structures or code.
\end{itemize}

For natural language understanding, we compared four possible frameworks (Microsoft's LUIS, Google's Dialogflow, Facebook's wit.ai and IBM's Watson) regarding important criteria for prototype implementation in a first step just for prototype 1. The comparison was extended for the frameworks moni.ai and Kauz.net for prototype 2 in the hackathon. All six frameworks support textual input and output, this was amongst others a basic requirement, but not all support complex conversation flows for advanced use cases. A comparison table for these criteria is shown in Table~\ref{tab:comparison}. As a result of the comparison, Google Dialogflow was chosen as a basic framework for prototype 1 based on the fulfillment of all requirements of prototype 1, one of which was the free availability. For prototype 2 and the hackathon, Google Dialogflow and IBM Watson Assistant were chosen, as an important factor next to the available user interface to enable non-programmers to work with the software was the fact that the providers agreed to accompany the hackathon event by sending experts for local support. 

\begin{table}[h]
  \centering
  \caption{Comparison of Microsoft's LUIS, Google's Dialogflow, Facebook's
    wit.ai, and IBM's Watson (from the requirements for the technical prototype (1) as in the original paper \cite{koetter18}, based on~\cite{nlucomparison}) and extended for prototype 2 by additional hackathon requirements and two new providers nameley moni.ai and Kauz.net (n.c. stands for Not Considered anymore or not yet for the prototype 1 or 2 as of early in 2018) } \centering
  \label{tab:comparison}
	\small
  \begin{tabular}{llllllll}
     \begin{sideways}\textbf{Requirement}\end{sideways}& & \begin{sideways}\textbf{LUIS}\end{sideways} &  \begin{sideways}\textbf{Dialogflow}\end{sideways} & \begin{sideways}\textbf{Wit.ai}\end{sideways} & \begin{sideways}\textbf{Watson}\end{sideways} & \begin{sideways}\textbf{moni.ai}\end{sideways} & \begin{sideways}\textbf{Kauz.net}\end{sideways} \\
    Overall (1,2) & Textual in-/output                     	& yes             & yes             & yes          & yes  & yes & yes       \\
    Overall (1,2) & German language                          & yes            & yes                                              & in Beta         & yes       & yes & yes \\
    Technical (1) & Python bindings                          & no             & yes & yes           & yes & n.c. & n.c.\\
    Technical (1) & Free service                             & no             & yes                                              & yes             & partly\footnote{10 000 free messages per month}      & n.c. & n.c.   \\
    Technical (1) & Remember state & yes            & yes                                              & yes             & yes        & n.c. & n.c.\\
    Technical (1) & Service bound                            & yes            & yes                                              & yes             & yes   & n.c. & n.c.     \\
    Technical (1) & Simple training                          & partly         & yes                                              & yes             & yes    & n.c. & n.c.   \\
    Hackathon (2) & Complex conversation flows           	& no              & yes             & n.c.    & yes  & n.c. & n.c.  \\
    Hackathon (2) & Provider support                        & n.c.      & yes             & n.c.   & yes  & no & yes      \\
    Hackathon (2) & User Interface  & yes & yes & n.c. & yes & yes & no
  \end{tabular}
\end{table}

\subsection{Prototype 1: claim management with technological extensions}
\label{prototype_tech}

\begin{figure*}[htb]
	\includegraphics[width=1.0\textwidth]{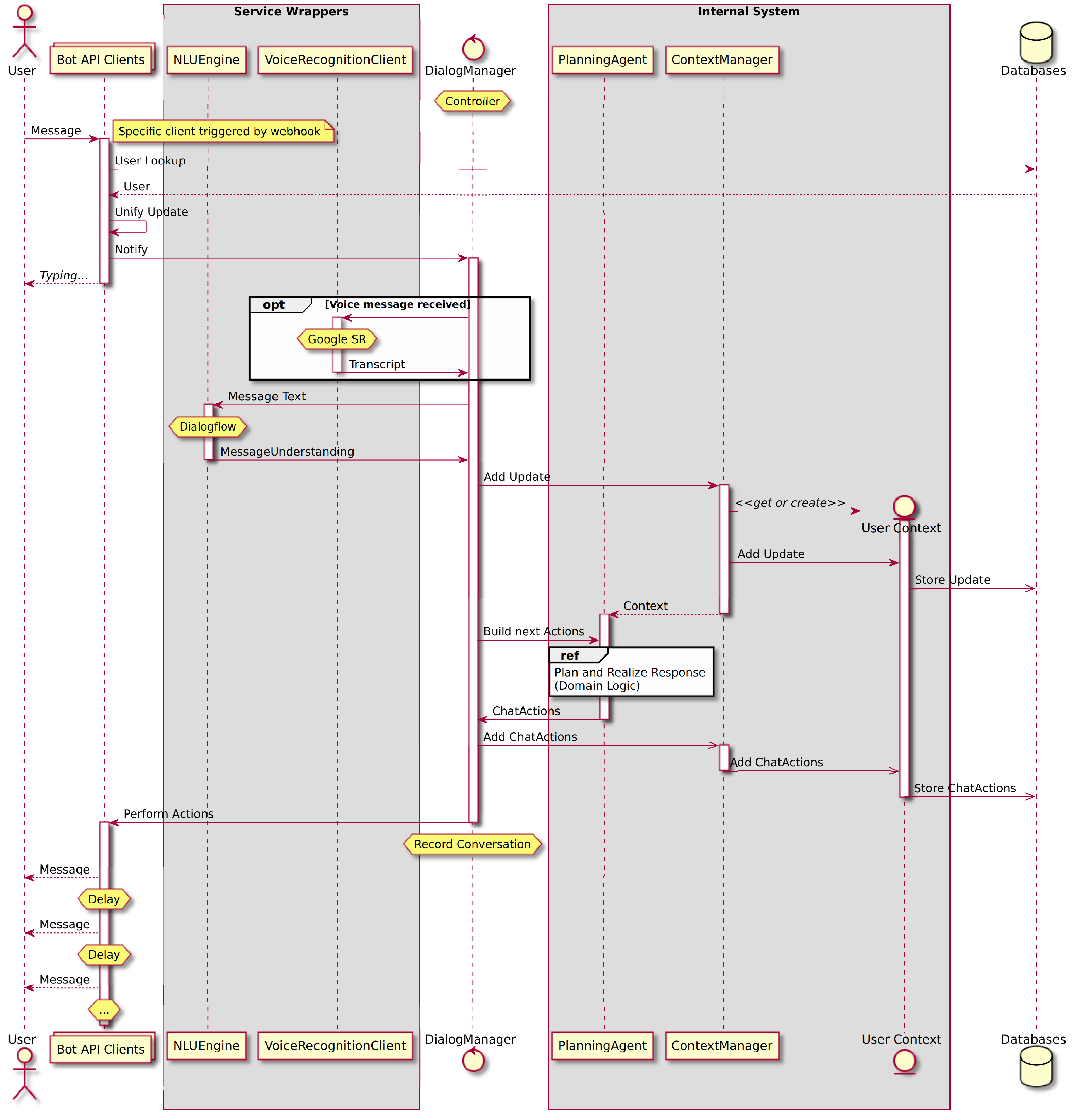}
	\centering
	\caption{Sequence diagram of the conversational agent prototype from \cite{koetter18}}
	\label{fig:sequencecore}
\end{figure*}

Prototype 1 fulfills the following scenario: \emph{The user has a damaged smartphone or tablet and wants to make an insurance claim}. The goal here is to focus on technology and build a demonstratable prototype in a 'traditional' project setting. We describe the results technically in the following. Figure~\ref{fig:sequencecore} shows the main components of the prototype and their operating sequence when processing a user message. To provide extensibility prototype architecture strictly separates service integration, internal logic and domain logic.

\begin{figure}
    \includegraphics[width=0.9\textwidth]{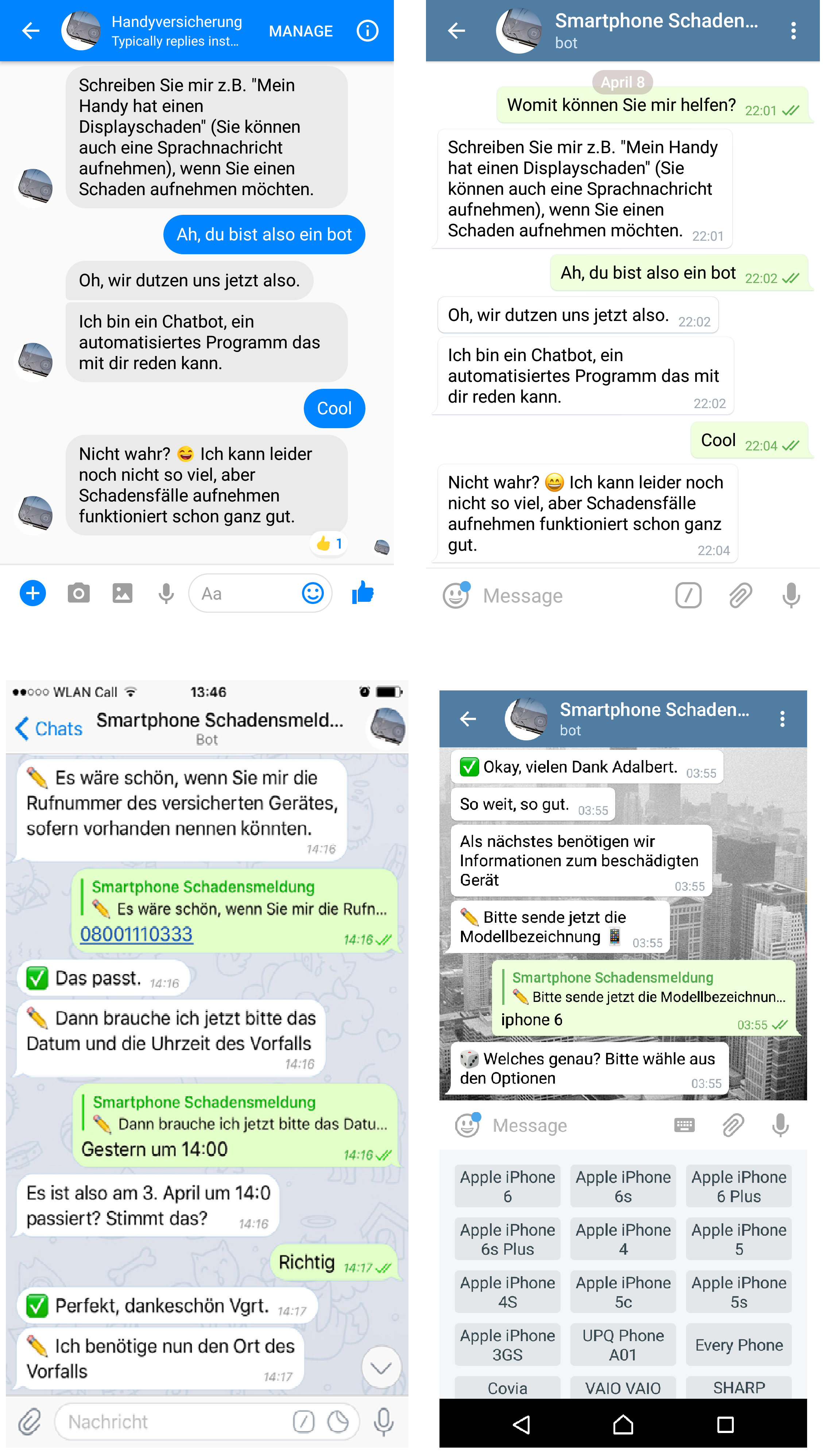}
  \caption{Top: The system is mirrored on both the Facebook and Telegram Messengers. Bottom left: Additional view with customer data input and intelligent recognition of words like \emph{yesterday}. Bottom right: Dialog excerpt of the prototype, showing the possibility to clarify the phone model via multiple-choice input. Extended version of figures in \cite{koetter18}.}
  \label{fig:demo}
\end{figure}

The user can interact with the bot over different communication channels which are integrated with different \emph{bot API clients}. To integrate a different messaging service, a new bot API client needs to be written. The remainder of the prototype can be reused. See Figure~\ref{fig:demo} for an example of the prototype on different communication channels.

Once a user has written a message, a lookup of \emph{user context} is performed to determine if a conversation with that user is already in progress. User context is stored in a database so no state is kept within external messaging services. Afterwards, a \emph{typing} notification is given to the user, indicating the bot has received the message and is working on it. This prevents multiple messages by a user who thinks the bot is not responsive.

In the next step, the message has to be understood by the bot. In case of a voice message, it is transcribed to text using a \emph{Google} speech recognition web service. Dialogflow is used for intent identification, which determines the function of a message and based on that a set of possible parameters~\cite{theconversationalinterface}. For example, the intent of the message ``the display of my smartphone broke'' may have the intent \texttt{phone\_broken} with the parameter \texttt{damage\_type} as \texttt{display\_damage}, while the parameter \texttt{phone\_type} is not given. Together, this information given by Dialogflow is a \texttt{MessageUnderstanding}

\begin{figure*}[htb]
	\includegraphics[width=1.0\textwidth]{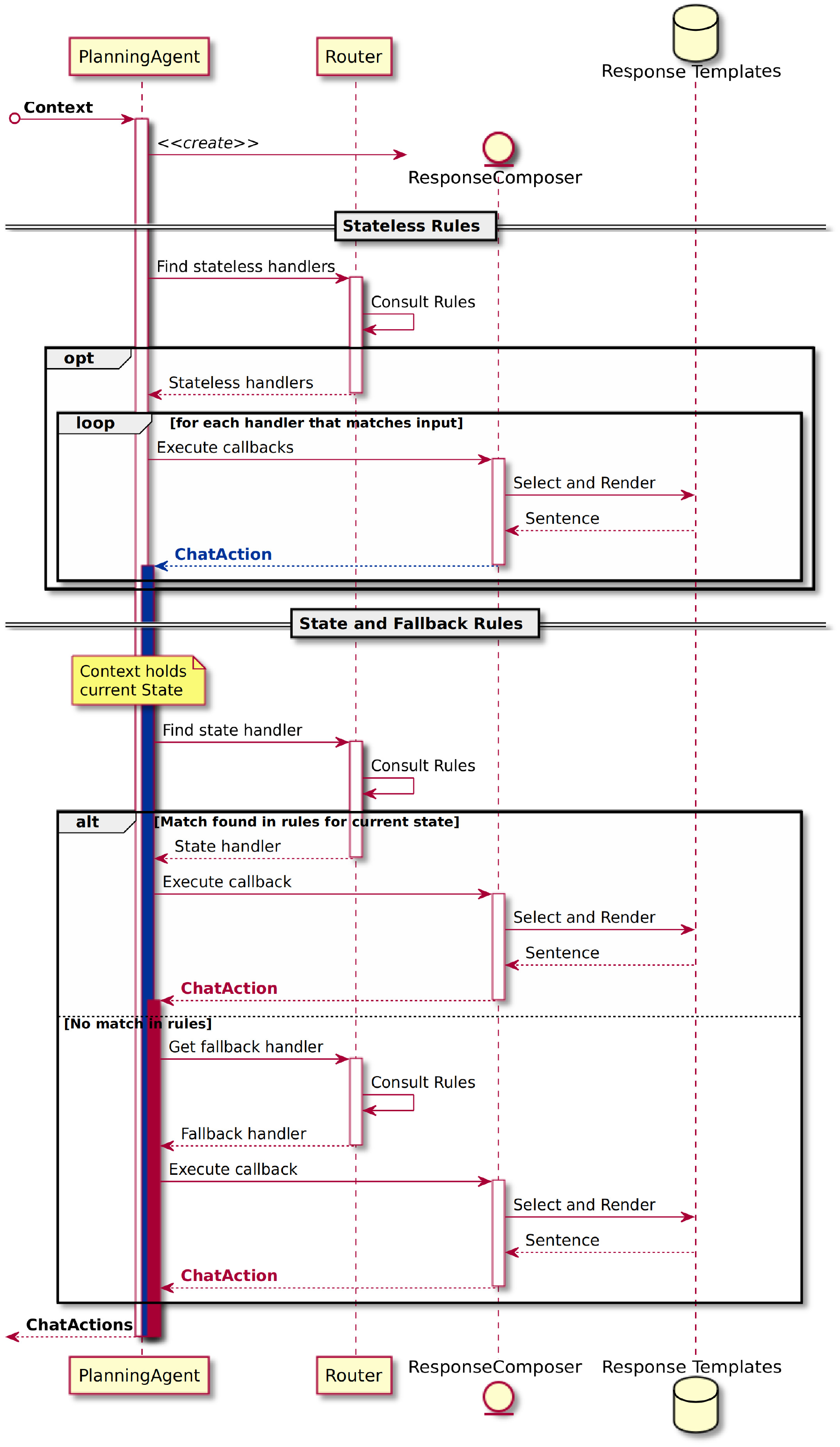}
	\centering
	\caption{Detailed sequence diagram of the response generation in the conversational agent prototype from \cite{koetter18}}
	\label{fig:sequencedomain}
\end{figure*}

As soon as the message is understood, the user context is updated. Afterwards, a response needs to be generated. This process, which was labeled with \emph{Plan and Realize Response} in Figure~\ref{fig:sequencecore}, is shown in detail in Figure~\ref{fig:sequencedomain}.

In the prototype, an agent-based strategy was chosen in order to combine the capabilities of the frame-based entities and parameters in Dialogflow with a custom dialog controller based on predefined rules in a finite state machine. This machine allows to define rules that trigger \emph{handlers} and \emph{state transitions} when a specific intent or entity-parameter combination is encountered. That way, both intent and frame processing happen in the same logically encapsulated unit, enabling better maintainability and extensibility. The rules are instances of a set of \texttt{*Handler} classes such as an \texttt{IntentHandler} for the aforementioned intent and parameter matching, supplemented by other handlers, e.g. an \texttt{AffirmationHandler}, which consolidates different intents that all express a confirmation along the lines of ``yes'', ``okay'', ``good'' and ``correct'', as well as a \texttt{NegationHandler}, a \texttt{MediaHandler} and an \texttt{EmojiSentimentHandler} (to analyze positive, neutral, or negative sentiment of a message with emojis). Each implements their own~ \texttt{matches(MessageUnderstanding)} method.

The following types of rules (handlers) are used within the dialog state machine:

\begin{enumerate}
\item \emph{Stateless} handlers are checked independently of the current state. For example, a \texttt{RegexHandler} rule determines whether the formality of the address towards the user should be changed (German differentiates the informal ``du'' and the formal ``Sie'')
\item \emph{Dialog States} map each possible state to a list of handlers that are applicable in that state. For instance, when the user has given an answer and the system asks for \emph{explicit confirmation} in a state \texttt{USER\_CONFIRMING\_ANSWER}, then an \texttt{AffirmationHandler} and a \texttt{NegationHandler} capture ``yes'' and ``no'' answers.
\item \label{itm:fallbacks} \emph{Fallback} handlers are checked if none of the applicable state handlers have yielded a match for an incoming \texttt{MessageUnderstanding}. These \texttt{fallbacks} include static, predefined responses with lowest priority (e.g. small talk), as well as handlers to repair the conversation by bringing the user back on track or changing the topic.
\end{enumerate}

At first, the system had only allowed a single state to be declared at the same time in the router. However, this had quickly proven to be insufficient as users are likely to want to respond or refer not only to the most recent message, but also to previous ones in the chat. With only a single contemporaneous state, the user's next utterance is always interpreted only in that state. In order to make this model resilient, every state would need to incorporate every utterance that the user is likely to say in that context. As this is not feasible, the prototype has state handlers that allow layering transitions on top of each other, allowing multiple simultaneous states which may advance individually.

To avoid an explosion of active states, the system has \emph{state lifetimes}: new states returned by callbacks may have a lifetime that determines the number of dialog moves this state is valid for. On receiving a new message, the planning agent decreases the lifetimes of all current dialog states by one, except for the case of utter non-understanding (``fallback'' intent). If a state has exceeded its lifetime, it is removed from the priority queue of current dialog states. 

Figure~\ref{fig:sequencedomain} contains details about how the system creates responses to user queries. Based on the applicable rule, the conversational agent performs chat actions (e.g. sending a message), which are generated from response templates, taking into account dialog state, intent parameters, and information like a user's name, mood and preferred level of formality. 

RuleHandlers, states and other dialog specific implementations are encapsulated, so a new type of dialog can be implemented without needing to change the other parts of the system.

Generated chat actions are stored in the user context and performed for the user's specific messenger using the \emph{bot API}. As the user context has been updated, the next message by the user continues the conversation.

The prototype explains its functionality and offers limited small talk. As soon as the user wants to make a damage claim, a predetermined questionnaire is used about type of damage, damaged phone, phone number, IMEI, damage time, damage event details, etc. Interpretation results of answers have to be confirmed by the user. For specific questions domain specific actions for clarification are implemented (see bottom right in Figure~\ref{fig:demo}). In a real-life application, claim management systems would be integrated to automatically trigger subsequent processes.

\subsection{Prototype 2: customer service and cross-selling in hackathon}
\label{prototype_customer}

For receiving more insights about the practicability of introducing chatbots to the insurance domain and for gaining experience with the usage of conversational frameworks, Fraunhofer IAO organized a four-day hackathon with five German insurance companies participating~\cite{kasper18}. This results in prototype 2 for the given task: \emph{Create a chatbot using IBM Watson Assistant or Google Dialogflow for answering questions about the annual bill of car insurance and leveraging cross-selling opportunities}. 

\begin{figure}
  \centering
    \includegraphics[width=0.9\textwidth]{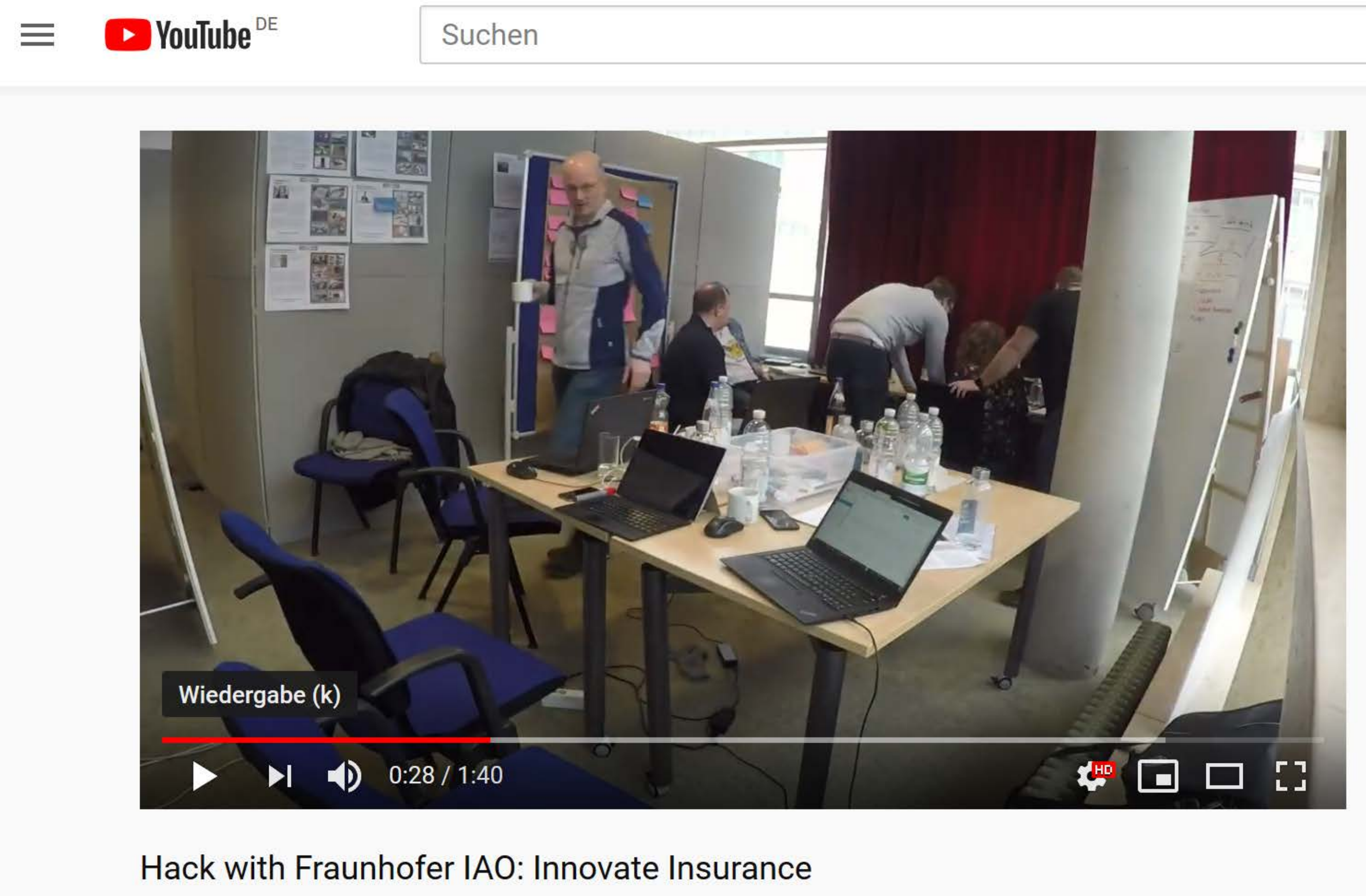}
  \caption{Hackathon impressions (www.youtube.com/watch?v=yHRLYJ\_olZ8, \cite{kasper18}).}
  \label{fig:hackathon}
\end{figure}

In the scope of the event four minimal products were created by four interdisciplinary teams of IT specialists, sales experts and other employees of the insurance companies. Doing so, in contrast to posing the challenge to external developers, our insurance partners were directly involved and could profit from the lessons learned from this \emph{internal hackathon}~\cite{hackathon14}. One impression from the resulting video is shown in Figure \ref{fig:hackathon} and also described in a blog article \cite{kasper18}.

Following four prototypes can be characterized as the teams worked independently: 
\begin{description} 
\item [Prototype A voice focus] One more technically oriented team started by adding voice technology to the chatbot for output purposes. Analogously, voice input could be used - although findings in the insurance project show that the input direction is more difficult to handle than the output direction. It showed that the focus is very entertaining in presentation and that the presentation especially of voice technology has to be performed carefully. In addition, the chatbot has been made more human-like by adding personal opinions on sports.
\item [Prototype B multimedia focus] Team B integrated several resources for better multimedia presentation, like images, videos, and the like. This already started with using an QR-code for accessing the prototype. The idea of using sophisticated multimedia content for explaining the annual bill like clickable graphics with videos has impressed the jury. The team focused therefore on customer experience and fine-tuned their interaction patterns by introducing delay in the response times.
\item [Prototype C stability and scope focus] The C-team focused on building a stable prototype for the complete task, achieving a large coverage of topics. The team spent most time in designing entities and intents as well as dialogue flow. This led to a comprehensive design and the most resilient result. The team was successful in maintaining background knowledge in a database and integrate it into the conversation flow.
\item [Prototype D customer identification focus] The team D focused on solving the customer identification issue. Using this information, they could give very detailed information on the current contract of the customer and the bill and use customer specific information for guiding the conversation itself. Another demonstration was the change of customer data. Additionally, some small talk was introduced for entertainment purposes.
\end{description}

Concerning the results, it is worth mentioning that all four groups succeeded in creating a usable product within the given timeframe that was able to handle the required use case of answering questions about annual bills for a small set of predefined queries. However, when letting a chatbot talk to people of other groups who were not involved in its development and thus not aware of the underlying dialog structure, the solutions proved to be error prone since they could not handle these unexpected inputs. This stems from \emph{hardcoding} parts of the scenario due to time constraints. In some cases, more expertise in dialog design would have helped anticipate typical user inputs. All teams worked with the \emph{entities} and \emph{intents} as are defined in most chatbot technologies, adding no programmable extensions (as compared to prototype 1). Therefore, the dialogue structure is static and creating the chatbot is done by adding intents and entities as well was \emph{if-then} like programming of the dialogue. The teams got only the plain frameworks of the providers and no specific extensions as for example for thorough testing. Working in teams on a chatbot proved to be helpful, but added organizational complexity, since resource access had to be shared and organized. Additionally, it turned out that professional content designers that build appropriate conversation models may be even more important than programmers, at least in case the used technology is enhanced enough.  A final survey among the participants showed that they enjoyed working in the chosen hackathon format and could benefit a lot from its results and lessons learned. 

\section{\uppercase{Evaluation}}
\label{sec:evaluation}

Both prototypes were evaluated using appropriate methods. For prototype one, a questionnaire-based approach with 14 participants was chosen. For prototype two, an expert comission had to choose and rate all four prototypes based on a very short questionnaire and come up with a point rating in a very short time span, as is typical for hackathons. We will first describe the results of both evaluation processes and then compare the results in the overall conclusion of this paper (see section \ref{sec:futurework}). 

\subsection{Evaluation of prototype 1 claim management}
\label{subsec:evaluation1}

To evaluate the produced prototype's quality and performance, we conducted a model trial with the goal to report a claim by using the chatbot without having any further instructions available.

Of the 14 participants (who all had some technical background), 35.7\% claimed to regularly use chatbots, 57.1\% to use them occasionally, and only 7.1\% stated that they had never talked to a chatbot before. However, all participants were able to report a claim within a range of about four minutes, resulting in an overall task completion rate of 100\%.

\begin{figure}[hpbt]
\centering
\begin{tikzpicture}[scale=0.85]
  \begin{axis}[
  xscale=0.8,
    ybar,
    enlargelimits=0.15,
    legend style={at={(0.5,-0.2)}, 
      anchor=north,legend columns=-1, ymin=0,ymax=10},
    ylabel={Average rating points},
    symbolic x coords={Ease of Use,Appropriate Formality,Natural Interaction,Response Quality,Personality,Funny \& Interesting, Entertainment, Lack of Deceptiveness
		},
    xtick=data,
    nodes near coords, 
	nodes near coords align={vertical},
    x tick label style={rotate=45,anchor=east},
    ]
    \addplot coordinates {(Ease of Use,8.0) (Appropriate Formality,8.3) 
		(Natural Interaction,7.9) (Response Quality,7.6) (Personality,5.2) (Funny \& Interesting,7.2) (Entertainment,7.7) (Lack of Deceptiveness,9.6) 
		};
  \end{axis}
\end{tikzpicture}
\caption{Survey results for prototype 1: average user experience ratings (fourteen participants, 0..10 points) from \cite{koetter18}.}\label{fig:eval}
\end{figure}
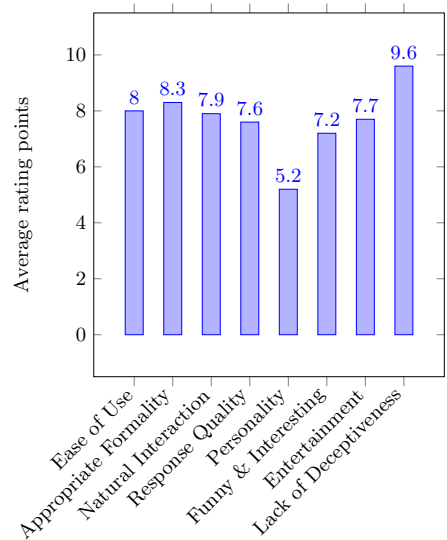

Additionally, the users had to rate the quality of their experiences with the conversational agent by filling out a questionnaire. For each question they could assign points between 0 (did not apply at all) and 10 (did apply to the full extent). The most important quality criteria, whose choice was oriented on the work of~\cite{evaluating2017}, are listed with their average ratings in Figure \ref{fig:eval} and are discussed in detail.

\begin{description}
\item[Ease of Use] With an average of 8 points for \emph{Ease of Use}, the users had no problems with using the bot to solve the task, since none of them gave less than 5 points. However, a variance of 2.46 still indicates a strong gap among the participants' experienced degree of usability.
\item[Appropriate Formality] 8.3 points on average for \emph{Appropriate Formality} indicate that the participants were comfortable with the formal and informal language the bot talked to them. Nonetheless, this criteria was also rated with points of only one and two. One of these users stated that he felt worried about permanently being called by his first name after he told it. Therefore, development of a more fine-grained detection mechanism for formal and informal language sould be considered in future versions of the chatbot, since for now we only rely on simple regular expressions.
\item[Natural Interaction] The rating for convincing \emph{Natural Interaction} with 7.9 points may be due to the fact that the conversation was designed in a strongly questionnaire-oriented way, which might have restricted the feeling of having a free user conversation. Therefore, improving the flexibility of the conversational flow and granting more freedom for user-centric dialog control might strengthen the authentic feeling during interaction with the agent.
\item[Response Quality] The satisfaction with given answers to users' domain specific questions was considered quite (but not totally) convincing with 7.6 points. Note that the high number of points might not be justified entirely, because the chatbot's implemented ability to answer questions is still very basic and restricted to concerns of claim handling. But, since the whole conversation is strongly driven by the agent itself, the users probably didn't find the time to ask many questions that went beyond the current limits of understanding. Connecting any kinds of knowledge bases might serve as a first future step towards extending the agent's response qualities.
\item[Personality] The least convincing experience was that chatbot's \emph{Personality}, which was rated with only 5.2 points on average. This is not surprising, since during this work we put comparatively less efforts in strengthening the agent's personal skills as it does not even introduce itself with a name, but instead mainly acts on a professional level, always concentrating on the fulfillment of its task. Facing these facts, a professional copywriter should have no problems developing a more convincing character for the chatbot.
\item[Funny \& Interesting] With 7.2 points, talking to the chatbot was experienced as quite \emph{Funny \& Interesting}, but still with a lot of room for further improvement. Again, the key here stays to loosen the strict procedure of forcing the user to finish the process and to allow more room for smalltalk and off topic contents.
\item[Entertainment] The agent's \emph{Entertainment} capabilities, which are at 7.7 points on average, could be upgraded by extending the conversational contents with additional enjoyable features not related to the questionnaire. At the moment, the chatbot is only able to tell some jokes from the insurance domain, but does not provide a holistic concept for customer entertainment.
\item[No Deception Feeling] The agent's \emph{lack of deceptiveness}, i.e. the degree to which users know it is not human, which at 9.6 points show that the bot's statements made its nature clear to users.
\end{description}

\subsection{Evaluation of prototype 2 as in the hackathon}

Typically, the evaluation in hackathons is done by a very short demo and questions from the audience by an expert committee. After four days, the resulting prototypes were examined by a jury considering the following predefined criteria. These differ strongly from the criteria in the first evaluation due to the time-constrained focused question. Most of the questions tackle the aforementioned \emph{response quality}, with a second thought on \emph{natural interaction}. They clarify what is actually a focus of the hackathon and what is not, putting emphasis on getting done. The \emph{formality} is just a subcriterion if the language style is adapted. All further going criteria like personality, fun, entertainment etc. have explicitly not been stated, but it is interesting that all groups put a strong focus on this during their presentation, trying to stand out from the field and enjoying to add human-like behavior. 

\begin{description}
	\item { \textbf{Language Support}
	\begin{itemize}
		\item Is the chatbot able to recognize the user input?
		\item Are the outputs of the chatbot adequate and understandable?
		\item Is the language style used by the bot adequate and consistent?
		\item Is the language style used by the bot adapted individually to user properties?
	\end{itemize}}
	\item { \textbf{Flexibility}
	\begin{itemize}
		\item Is the chatbot able to correctly recognize input even in unusual phrases?
		\item Is the chatbot able to respond appropriately to unexpected input?
	\end{itemize}}
	\item { \textbf{Scope of Functions}
	\begin{itemize}
		\item How well has the scenario \emph{Annual fee bill} been covered by the prototype?
		\item How well are extensions and transitions implemented leading to other topics such as cross selling?
	\end{itemize}}
	\item { \textbf{Presentation}
	\begin{itemize}
		\item Is the presentation convincing? Are there any differences to the other groups?
		\item Did the team manage to explain the chatbot in a timely manner?
	\end{itemize}}
\end{description}

The expert committee had difficulties finding a winner in the given time span, as all of the four prototypes fulfilled certain aspects that were identified as interesting to the jury. Altogether the participants and experts learned a lot about chatbot design and technologies and were satisfied with the results. Both technologies led to good results in the hackathon, but most of the participants felt they needed provider support and expert knowledge on dialogue design for creating a real product for customers. More overall conclusions follow in the next section.

\section{\uppercase{Conclusions and Outlook}}
\label{sec:futurework}

In this work we have shown how conversational agents can be applied for different use cases in theory and practice. We showed how our classification of conversation agents applies for the prototypes generated for our two scenarios: (1) claim management process support in traditional project setting as well as (2) customer service and cross selling in an interdisciplinary hackathon. The potential processes to employ chatbots have been shown in general for the insurance companies and focused on customer service processes. One key result of our former paper \cite{koetter18} containing  prototype 1 is a system of multiple conversational states enabling more flexible conversations. We extended the evaluation with real users and additionally showed the prototype to various customer groups, small businesses and more insurance companies. Altogether, the prototype is able to handle the scenario satisfactory. One possible improvement is the point of \emph{realism}, for example by more human-like behavior in a consistent persona and better determination of the desired degree of formality. The newly performed and described activity is the creation of prototype 2 for customer service consisting of actually four prototypes in an interdisciplinary hackathon. It has shown that the results here differ strongly form prototype 1 due to different goals, different time span and different skills. No extensions to the entity-intent concept were performed, but several innovative ideas have been included like multimedia integration, voice integration, several entertaining aspects and especially persona design. Prototype 1 did not have a name in the beginning, whereas all teams came up with innovative names for prototype 2 at the beginning of the hackathon. 

Altogether, after two years of chatbot experience, we can summarize the potentials for the conversational agent technology: 
\begin{description}
\item [Maturity of technology] Technology matures and is more often perceived in all day activities, most people know chatbots and how they can be used. Many people aready have already tried out a chatbot.
\item [Service enhandement] Agents can be used for better availability (24/7/365) and to reduce the workload of the service staff. 
\item [Tools] Tools are available especially for good English language support. More languages and features are added as time passes. New technological frameworks are available, the existing ones are improved. 
\item [Simple tasks] Easy application for simple tasks and simple prototype creation is possible in a short time span. Transfer from prototype to live system is still more difficult. 
\item [Applicability] Many application scenarios are possible. 
\end{description}

We have identified following challenges for new conversational agents and especially for transitioning from demo to prototype to live service:

\begin{description}
\item[Testing] is time-consuming and error-prone
\item[Domain language] has to be usually hardcoded or added manually supported by machine learning in an optimal case
\item[Handovers] Designing handovers is a challenging tasks that not all frameworks fulfill perfectly 
\item[Maintenance] of the chatbot and further development is a challenging tasks and the process has to be defined
\item[Self-learning] is not available in the expected scope as lots of people have very high and unrealistic expectations to machine learning 
\item[High expectations] to chatbot technology in general which might not be fulfilled in the beginnning
\item[Security and integration] issues as with most technologies
\end{description} 

As a result of our two prototypes, the evaluations and the hackathon participant surveys, we came up with the following success factors that we believe need to be respected when planning to introduce conversational agents to companies:

\begin{description}
	\item [Clear scope definition] Use cases and functionality of the conversational agent should be predefined as detailed as possible.
	\item [Customer-oriented development] Tests with intended audience and changing test participants to prevent them getting used to the dialog structure.
	\item[Careful improvements and testing] Sufficient time and care should be invested in testing and improving the agent. A nonfunctional or only partly functional bot deployed to the public too early might cause a negative reception that cannot be corrected with future improvements.
	\item[Perform regression tests] Especially for self-learning agents it is crucial to ensure that the bot does not "`unlearn"' skills that once worked successfully.
	\item[Facilitate maintenance] Provide high-level (graphical) dialog customization options for the employees of the related department for supporting easy extension and improvement of the agent.
	\item[Choice of technology provider] Technology providers should be compared and chosen according to the company environment and its conditions. One partner should be selected for longer cooperation.
\end{description}
	
The model trial covering two prototypes has shown that conversational agents are ready for productive use. However, the effort in creating and maintaining a conversational agent is not to be underestimated. While a successful conversation with a chatbot provides a satisfying customer experience, errors and gaps in dialog flow let user satisfaction drop rather quickly. While users do not expect a human like conversation and phrase their statements accordingly, they expect clearly formulated requests and answers to be readily understood. Currently we are working on supporting small and medium enterprises with evaluation of the technology and potential use cases for their businesses. In future research, we would like to implement a real-life conversational agent as well as perform a real-life evaluation with an insurance partner to quantify the benefits of agent use, e.g. call reduction, success rate, and customer satisfaction as well as support small and medium businesses with agent creation.

\section*{Acknowledgements}
The work is partially based on work carried out in the project 'Business Innovation Engineering Center', which is funded by the Ministry of Economic Affairs, Labour and Housing Baden-Wuerttemberg under the reference number 3-4332.62-IAO/56. The hackathon itself was conducted within the 'Innovationsnetzwerk Digitalisierung fuer Versicherungen' \cite{iao2018innonetz}. The authors want to thank all participants for their contributions and feedback. 

%
%
\bibliographystyle{splncs04}
\bibliography{./bibliography/bibliography}

\end{document}